\documentclass[preprint,preprintnumbers,amsmath,amssymb,floatfix,nofootinbib]{revtex4}
\usepackage{graphicx}
\usepackage{url}

\begin{document}          
\preprint{FSU-HEP-2006-0410}
\preprint{hep-ph/0606102}
\title{NLO QCD corrections to $W$ boson production with a massive
  $b$-quark jet pair at the Fermilab Tevatron $p\bar p$ collider}
\author{F.~Febres~Cordero}
\email{ffebres@hep.fsu.edu}
\affiliation{Physics Department, Florida State University,
Tallahassee, FL 32306-4350, USA}
\author{L.~Reina}
\email{reina@hep.fsu.edu}
\affiliation{Physics Department, Florida State University,
Tallahassee, FL 32306-4350, USA}
\author{D.~Wackeroth}
\email{dow@ubpheno.physics.buffalo.edu}
\affiliation{Department of Physics, SUNY at Buffalo,
Buffalo, NY 14260-1500, USA}

\date{\today}

\begin{abstract}

We calculate the Next-to-Leading Order (NLO) QCD corrections to $W
b\bar{b}$ production including full bottom-quark mass effects. We
study the impact of NLO QCD corrections on the total cross section and
invariant mass distribution of the bottom-quark jet pair at the
Fermilab Tevatron $p\bar p$ collider.  We perform a detailed
comparison with a calculation that considers massless bottom quarks.
We find that neglecting bottom-quark mass effects overestimates the
NLO total cross-section for $Wb\bar{b}$ production at the Tevatron by
about 8\% independent of the choice of renormalization and
factorization scale.
\end{abstract}
%
\maketitle

\section{Introduction}
\label{sec:intro}
The associated production of a $W$ boson with a $b\bar{b}$ pair plays
a critical role at the Fermilab Tevatron $p\bar{p}$ collider, since it
accounts for one of the most important background processes to both
the associated production of a Higgs boson with a $W$ boson,
$p\bar{p}\rightarrow HW$ (with $H\rightarrow
b\bar{b}$)~\cite{Abulencia:2005ep,CDFnote:2006Wbb,Abazov:2004jy,D0note:2006Wbb,Patwa:2006rd},
and single-top production, $p\bar{p}\rightarrow t\bar{b},\bar{t}b$
(with $t(\bar{t})\rightarrow W
b(\bar{b})$)~\cite{Acosta:2004bs,Abazov:2005zz,Gresele:2006se}. These
two processes are of extreme relevance to the physics program of the
Tevatron: they both test fundamental predictions of the Standard Model
(SM), i.e. the existence of a Higgs boson and the
structure of the $Wtb$ vertex, and at the same time constitute a
window to new physics. The cross section for $p\bar{p}\rightarrow HW$
has been calculated including up to Next-to-Next-to-Leading (NNLO) QCD
corrections~\cite{Han:1991ia,Mrenna:1997wp,Brein:2003wg} and
$O(\alpha)$ electroweak corrections~\cite{Ciccolini:2003jy}, while
single-top production has been calculated at Next-to-Leading (NLO) in
QCD~\cite{Stelzer:1997ns,Stelzer:1998ni,Smith:1996ij,Harris:2002md,Sullivan:2004ie,Cao:2004ky,Cao:2004ap,Cao:2005pq,Sullivan:2005ar},
and at one-loop of electroweak (SM and MSSM)
corrections~\cite{Beccaria:2006ir}.

The production of a Higgs boson in association with an electroweak
gauge boson, $p\bar p \to HV$ ($V\!=\!Z,W$) with $H\to b\bar{b}$, is
the most sensitive production channel of a SM Higgs boson at the
Tevatron for a Higgs boson lighter than about 140~GeV.  A relatively
light SM Higgs boson is preferred by electroweak precision data, $M_H
= 89^{+42}_{-30}$ GeV at 68\% confidence
level~\cite{lepewwg:2005di}~\footnote{For an update see the LEPEWWG
  website at http://lepewwg.web.cern.ch/LEPEWWG}.  The Tevatron with an
integrated luminosity of $6 \; {\rm fb}^{-1}$ will be able to exclude
a Higgs boson with $115 \; {\rm GeV}<M_H<180$ GeV at 95\% confidence
level~\cite{sonnenschein}, which will provide important guidance for
the search strategy at the CERN Large Hadron Collider.  Thus, to fully
exploit the Tevatron's potential to detect the SM Higgs boson or to
impose limits on its mass, it is crucial that the dominant background
processes are under good theoretical control.

In the present experimental analyses\footnote{For updated results,
see the CDF and D$\emptyset$ websites at
www-cdf.fnal.gov/physics/exotic/exotic.html and
www-d0.fnal.gov/Run2Physics/WWW/results/higgs.htm.} the effects of NLO
QCD corrections on the total cross-section and the dijet invariant
mass distribution of the $Wb\bar{b}$ background process have been
taken into account by using the MCFM package~\cite{MCFM:2004}. In
MCFM, the NLO QCD predictions of both total and differential
cross-sections for the $q\bar{q}^\prime\rightarrow Wb\bar{b}$
production process have been calculated in the zero bottom-quark mass
($m_b=0$) approximation~\cite{Ellis:1998fv,Campbell:2002tg}.  From a
study of the Leading Order (LO) cross-section, finite bottom-quark
mass effects are expected to affect both total and differential
$Wb\bar{b}$ cross-sections mostly in the region of small
$b\bar{b}$-pair invariant masses~\cite{Campbell:2002tg}. Given the
variety of experimental analyses involved both in the search for $HW$
associated production and single-top production, it is important to
precisely assess the impact of a finite bottom-quark mass over the
entire kinematical reach of the process, including complete NLO QCD
corrections.

In this paper we compute the NLO QCD corrections to
$q\bar{q}^\prime\rightarrow Wb\bar{b}$, including full bottom-quark
mass effects.  Using the MCFM package~\cite{MCFM:2004}, we compare our
results with the corresponding results obtained in the $m_b=0$ limit.
Numerical results are presented for the total cross-section and the
invariant mass distribution of the $b\bar{b}$ jet pair, for the
Tevatron $p\bar{p}$ collider, including kinematic cuts and a
jet-finding algorithm. In particular, we apply the $k_T$ jet algorithm
and require two tagged $b$-jets in the final state.

The outline of the paper is as follows.  In
Section~\ref{sec:calculation} we briefly discuss the technical details
of our calculation, while we present numerical results and a
discussion of the bottom-quark mass effects in
Section~\ref{sec:results}. Section~\ref{sec:conclusions} contains our
conclusions.

\section{Calculation}
\label{sec:calculation}
\begin{figure}[htb]
\begin{center}
\includegraphics[scale=0.75]{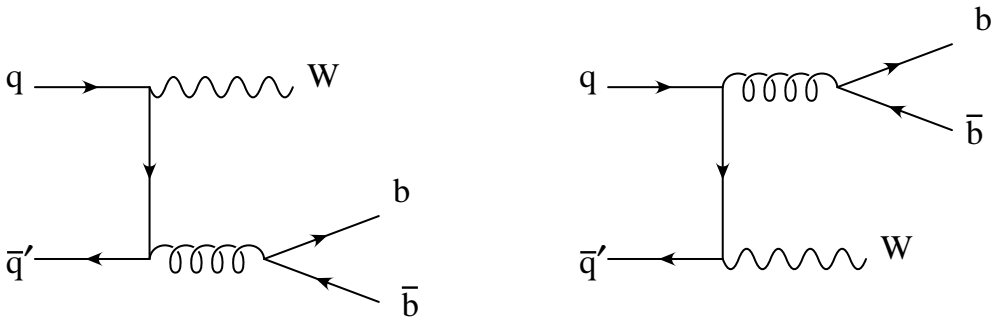} 
\caption[]{Tree level Feynman diagrams for $q\bar{q}^\prime\rightarrow b\bar{b}W$.}
\label{fig:tree_level}
\end{center}
\end{figure}

The NLO QCD corrections to $q\bar{q}^\prime\rightarrow b\bar{b}W$
consist of both one-loop virtual corrections to the tree level
processes depicted in Fig.~\ref{fig:tree_level} and one-parton real
radiation from both the initial and final state quarks,
i.e. $q\bar{q}^\prime\rightarrow b\bar{b}W+g$. At the same order, the
$qg(\bar{q}g)\rightarrow b\bar{b}W+q(\bar{q})$ process also needs to be
included. 

\begin{figure}[htb]
\begin{center}
\includegraphics[scale=0.55,angle=-90]{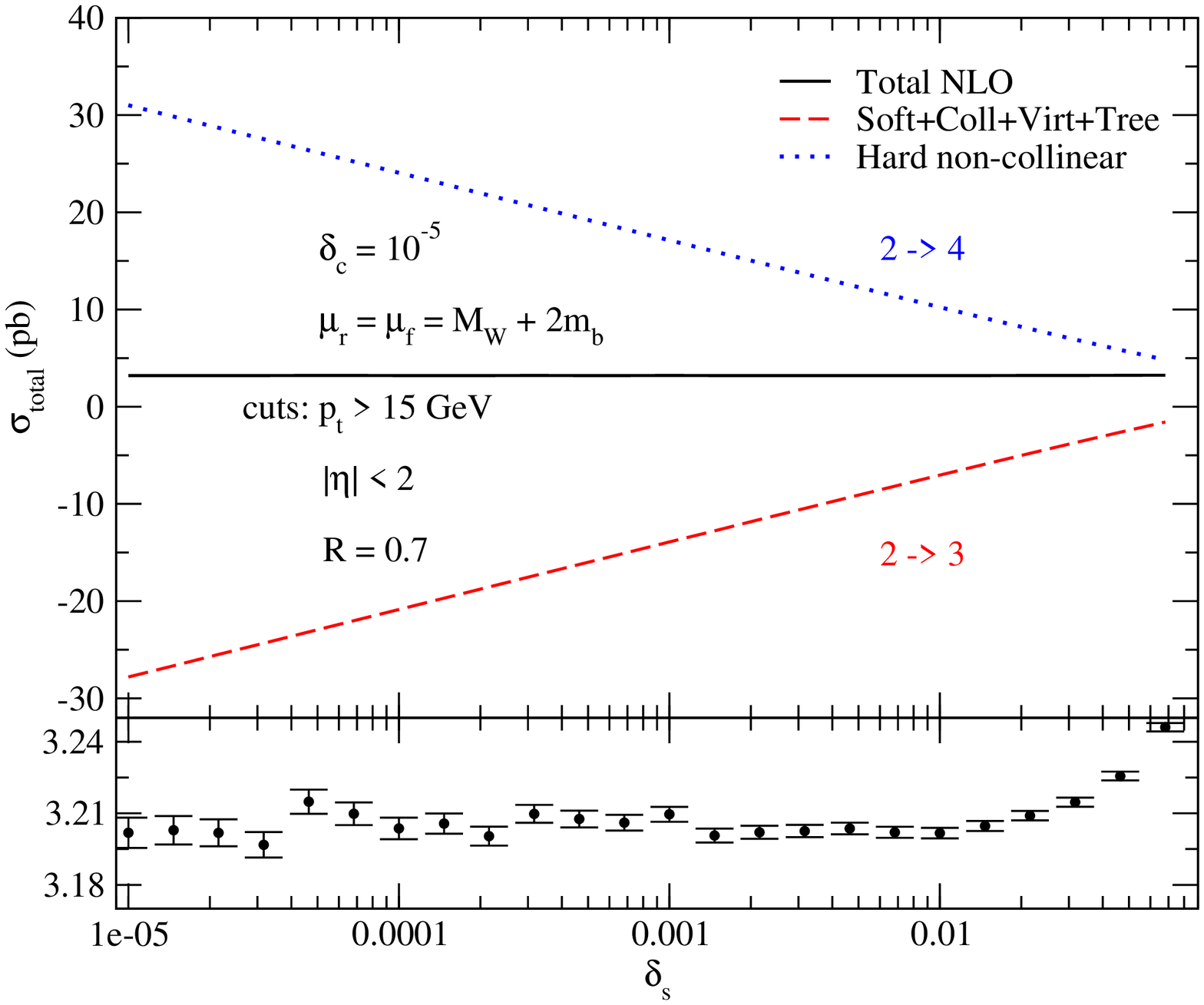} 
\caption[]{Dependence of the total NLO QCD cross-section on the
  $\delta_s$ PSS parameter, when $\delta_c$ is fixed at
  $\delta_c=10^{-5}$. In the upper window we illustrate separately the
  cutoff dependence of the soft and hard-collinear part ($2\rightarrow
  3$, red dashed curve) and of the hard non-collinear part
  ($2\rightarrow 4$, blue dotted curve) of the real corrections to the
  total cross-section.  The $2\rightarrow 3$ curve also includes those
  parts of the $2\rightarrow 3$ NLO cross-section that do not depend
  on $\delta_c$ and $\delta_s$, i.e. the tree level and one-loop
  virtual contributions.  The sum of all the contributions corresponds
  to the black solid line. The lower window shows a blow-up of the
  black solid line in the upper plot, to illustrate the stability of
  the result. The error bars indicate the statistical uncertainty of
  the Monte Carlo integration.}
\label{fig:ds_dependence}
\end{center}
\end{figure}
\begin{figure}[htb]
\begin{center}
\includegraphics[scale=0.55,angle=-90]{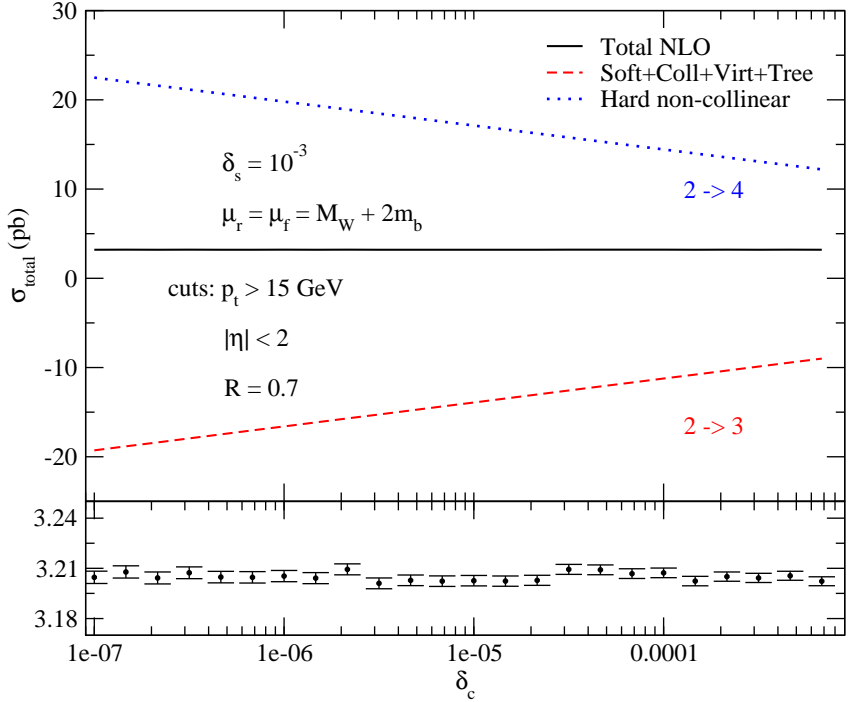} 
\caption[]{Dependence of the total NLO QCD cross-section on the
  $\delta_c$ PSS parameter, when $\delta_s$ is fixed at
  $\delta_s=10^{-3}$. In the upper window we illustrate separately the
  cutoff dependence of the soft and hard-collinear part ($2\rightarrow
  3$, red dashed curve) and of the hard non-collinear part
  ($2\rightarrow 4$, blue dotted curve) of the real corrections to the
  total cross-section. The $2\rightarrow 3$ curve also includes those
  parts of the $2\rightarrow 3$ NLO cross-section that do not depend
  on $\delta_c$ and $\delta_s$, i.e. the tree level and one-loop
  virtual contributions.  The sum of all the contributions corresponds
  to the black solid line. The lower window shows a blow-up of the
  black solid line in the upper plot, to illustrate the stability of
  the result. The error bars indicate the statistical uncertainty of
  the Monte Carlo integration.}
\label{fig:dc_dependence}
\end{center}
\end{figure}

The $O(\alpha_s)$ virtual corrections consist of self-energy, vertex,
box, and pentagon diagrams with several massive propagators, since we
take $m_b\neq 0$. They contain both UV and IR singularities which need
to be computed analytically. For box and pentagon diagrams we use
techniques similar to the ones explained in detail in
Refs.~\cite{Reina:2001bc,Dawson:2003zu}. We use dimensional
regularization with $d=4-2\epsilon$, and extract both UV and IR
divergences as poles in $\epsilon$. The UV singularities are cancelled
by introducing a suitable series of counterterms. We renormalize the
wave functions of the external quark fields in the on-shell scheme,
and the strong coupling constant $\alpha_s$ in the
$\overline{MS}$-scheme, decoupling the top quark. At this order in QCD
the weak vertex renormalization consists only of the external quark
wave-function renormalization. Self-energy, vertex, box, and pentagon
diagrams contain IR divergences that combine and cancel against the
analogous divergences in the real emission $O(\alpha_s)$ corrections,
and in the renormalized Parton Distribution Functions (PDF).

We compute the real emission $O(\alpha_s)$ corrections using the Phase
Space Slicing (PSS) method with two cutoffs: $\delta_s$ for the soft
singularities, and $\delta_c$ for the hard-collinear
singularities~\cite{Harris:2001sx,Reina:2001bc,Dawson:2003zu}. The
independence of the final result on the arbitrary values of $\delta_s$
and $\delta_c$ has been checked over a large range of values for both
parameters and is illustrated in Figs.~\ref{fig:ds_dependence} and
\ref{fig:dc_dependence}. The numerical results in
Section~\ref{sec:results} have been obtained using $\delta_s=10^{-3}$
and $\delta_c=10^{-5}$.

In our calculation we treat $\gamma_5$ according to the naive
dimensional regularization approach, i.e.  we enforce that $\gamma_5$
anticommutes with all other $\gamma$ matrices in $d=4-2\epsilon$
dimensions. This is known to give origin to inconsistencies when at
the same time the $d$-dimensional trace of four $\gamma$ matrices and
one $\gamma_5$ is forced to be non-zero (as in $d=4$, where
$Tr(\gamma^\mu\gamma^\nu\gamma^\rho\gamma^\sigma\gamma_5)=4i\epsilon^{\mu\nu\rho\sigma}$)
~\cite{Larin:1993tq}.  In our calculation both UV and IR divergences
are handled in such a way that we never have to enforce simultaneously
these two properties of the Dirac algebra in $d$ dimensions. For
instance, the UV divergences are extracted and cancelled at the
amplitude level, after which the $d\rightarrow 4$ limit is taken and
the renormalized amplitude is squared using $d=4$. Thus, all fermion
traces appearing at this point are computed in four dimensions and
therefore have no ambiguities.

Both virtual and real corrections have been checked by independent
calculations that have used FORM~\cite{Vermaseren:2000nd},
TRACER~\cite{Jamin:1991dp}, the FF
package~\cite{vanOldenborgh:1990yc}, and MAPLE. The $2\rightarrow 4$
amplitudes for the real corrections have been double checked using
Madgraph~\cite{Murayama:1992gi,Stelzer:1994ta,Maltoni:2002qb}.

\section{Numerical results}
\label{sec:results}
In this paper we present results for $Wb\bar{b}$ production at the
Tevatron, including NLO QCD corrections, and using a non-zero
bottom-quark mass, fixed at $m_b$=4.62 GeV. The $W$ boson is considered
on-shell and its mass is taken to be $M_W=80.41$~GeV.  The mass of the
top quark, entering in virtual corrections, is set to
$m_t=174$~GeV. The LO results use the 1-loop evolution of $\alpha_s$
and the CTEQ6L set of PDF~\cite{Lai:1999wy}, while the NLO results use
the 2-loop evolution of $\alpha_s$ and the CTEQ6M set of PDF, with
$\alpha_s^{NLO}(M_Z)=0.118$.  The $W$ boson coupling to quarks is
proportional to the Cabibbo-Kobayashi-Maskawa (CKM) matrix elements.
We take $V_{ud}=V_{cs}=0.975$ and $V_{us}=V_{cd}=0.222$, while we
neglect the contribution of the third generation, since it is
suppressed either by the initial state quark densities or by the
corresponding CKM matrix elements.

We implement the $k_T$ jet
algorithm~\cite{Catani:1992zp,Catani:1993hr,Ellis:1993tq,Kilgore:1996sq}
with a pseudo-cone size $R=0.7$ and we recombine the parton momenta
within a jet using the so called covariant
$E$-scheme~\cite{Catani:1993hr}. We checked that our implementation of
the $k_T$ jet algorithm coincides with the one in MCFM.  We require
all events to have a $b\bar{b}$ jet pair in the final state, with a
transverse momentum larger than $15$~GeV ($p_T^{b,\bar{b}}>15$~GeV)
and a pseudorapidity that satisfies $|\eta^{b,\bar{b}}|<2$. We impose
the same $p_T$ and $|\eta|$ cuts also on the extra jet that may arise
due to hard non-collinear real emission of a parton, i.e. in the
processes $Wb\bar{b}+g$ or $Wb\bar{b}+q(\bar{q})$. This hard
non-collinear extra parton is treated either \emph{inclusively} or
\emph{exclusively}, following the definition of \emph{inclusive} and
\emph{exclusive} as implemented in the MCFM code~\cite{MCFM:2004}.  In
the \emph{inclusive} case we include both two- and three-jet events,
while in the \emph{exclusive} case we require exactly two jets in the
event. Two-jet events consist of a bottom-quark jet pair that may also
include a final-state light parton (gluon or quark) due to the applied
recombination procedure. Results in the massless bottom-quark
approximation have been obtained using the MCFM code~\cite{MCFM:2004}.

\begin{figure}[htb]
\begin{center}
\includegraphics*[scale=0.5]{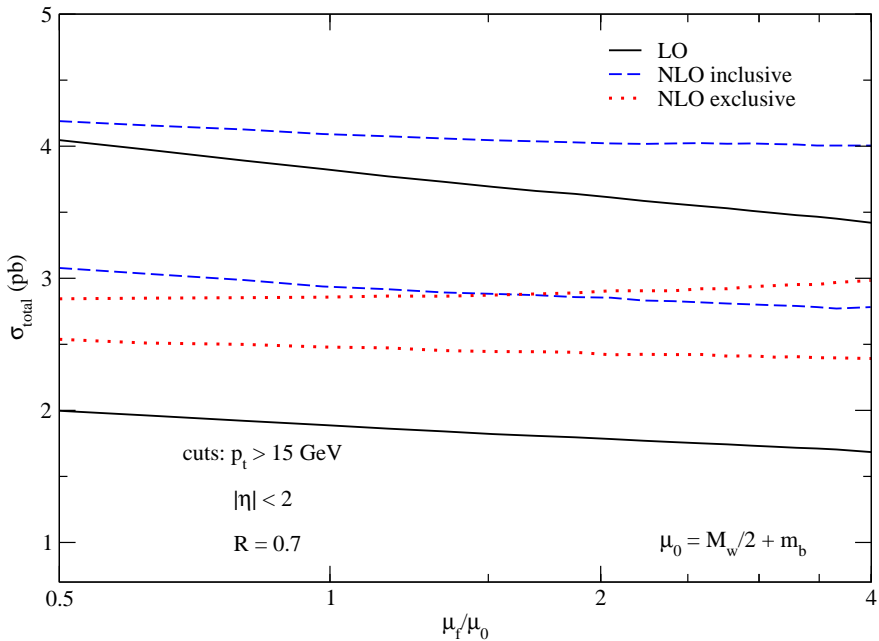} 
\caption[]{Dependence of the LO (black solid band), NLO
\emph{inclusive} (blue dashed band), and NLO
\emph{exclusive} (red dotted band) total cross-sections on the
renormalization/factorization scales, including full bottom-quark
mass effects. The bands are obtained by varying both $\mu_R$ and
$\mu_F$ between $\mu_0/2$ and $4\mu_0$ (with $\mu_0=m_b+M_W/2$).}
\label{fig:mu_dependence_band}
\end{center}
\end{figure}
\begin{figure}[htb]
\begin{center}
\includegraphics*[scale=0.6]{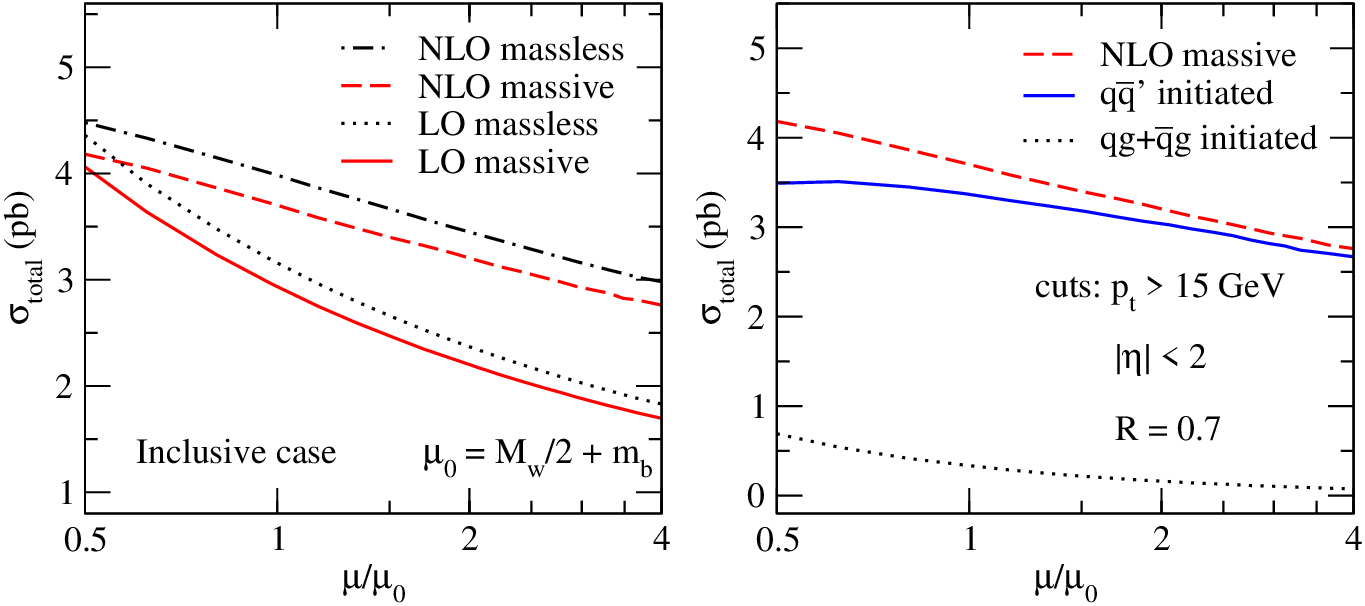} 
\caption[]{Dependence of the LO and NLO \emph{inclusive} total cross-section
on the renormalization/factorization scale, when
$\mu_R\!=\!\mu_F$. The left hand side plot compares both LO and NLO
total cross-sections for the case in which the bottom quark is treated
as massless (MCFM) or massive (our calculation).  The right hand side
plot shows separately, for the massive case only, the scale dependence
of the $q\bar{q}^\prime$ and $qg+\bar{q}g$ contributions, as well as
their sum.}
\label{fig:mu_dependence_inc}
\end{center}
\end{figure}

\begin{figure}[htb]
\begin{center}
\includegraphics*[scale=0.6]{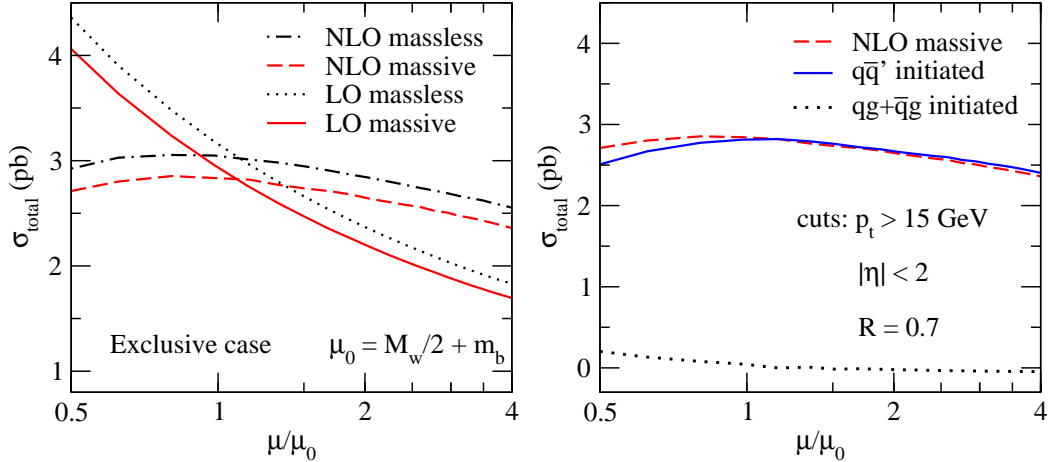} 
\caption[]{Dependence of the LO and NLO \emph{exclusive} total cross-section
on the renormalization/factorization scale, when
$\mu_R\!=\!\mu_F$. The left hand side plot compares both LO and NLO
total cross-sections for the case in which the bottom quark is treated
as massless (MCFM) or massive (our calculation).  The right hand side
plot shows separately, for the massive case only, the scale dependence
of the $q\bar{q}^\prime$ and $qg+\bar{q}g$ contributions, as well as
their sum.}
\label{fig:mu_dependence_exc}
\end{center}
\end{figure}

\begin{figure}[htb]
\begin{center}
\includegraphics*[scale=0.5]{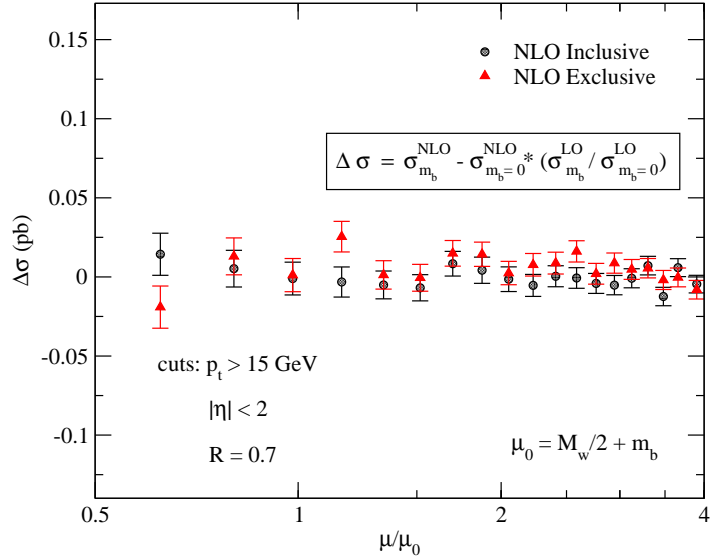} 
\caption[]{Dependence 
on the renormalization/factorization scale of the rescaled difference 
between our NLO calculation (with $m_b\ne 0$) of the total cross-section 
and MCFM (with
$m_b=0$) for the \emph{inclusive} and \emph{exclusive} cases
(with $\mu_R\!=\!\mu_F$). The error bars indicate the statistical uncertainty of
  the Monte Carlo integration.}
\label{fig:sigma_ratio_NLO}
\end{center}
\end{figure}

In Figs.~\ref{fig:mu_dependence_band}-\ref{fig:mu_dependence_exc} we
illustrate the renormalization and factorization scale dependence of
the LO and NLO total cross-sections, both in the \emph{inclusive} and
\emph{exclusive} case.  Fig.~\ref{fig:mu_dependence_band} shows the
overall scale dependence of both LO, NLO \emph{inclusive} and NLO
\emph{exclusive} total cross-sections, when both $\mu_R$ and $\mu_F$
are varied independently between $\mu_0/2$ and $4\mu_0$ (with
$\mu_0=m_b+M_W/2$), including full bottom-quark mass effects. We
notice that the NLO cross-sections have a reduced scale dependence
over most of the range of scales shown, and the \emph{exclusive} NLO
cross-section is more stable than the \emph{inclusive} one especially
at low scales.  This is consistent with the fact that the
\emph{inclusive} NLO cross-section integrates over the entire 
phase space of the $qg(\bar{q}g)\rightarrow b\bar{b}W + q(\bar{q})$
channels that are evaluated with NLO $\alpha_s$ and NLO PDF, but are
actually tree-level processes and retain therefore a strong scale
dependence. In the
\emph{exclusive} case only the $2\rightarrow 3$ collinear kinematic of
these processes is retained, since 3-jets events are discarded, and
this makes the overall renormalization and factorization scale
dependence milder.  To better illustrate this point, we show in the
right hand side plots of Figs.~\ref{fig:mu_dependence_inc} and
\ref{fig:mu_dependence_exc} the mu-dependence of the total cross
section and of the partial cross-sections corresponding to the
$q\bar{q}^\prime$ and the $qg+\bar{q}g$ initiated channels separately,
for $\mu_R=\mu_F$, both for the
\emph{inclusive} and for the \emph{exclusive} case.  It is clear that
the low scale instability of the \emph{inclusive} cross-section is
entirely driven by the $qg+\bar{q}g$ contribution.  In the left hand
side plots of Figs.~\ref{fig:mu_dependence_inc} and
\ref{fig:mu_dependence_exc} we also compare the scale dependence of
our results to the scale dependence of the corresponding results
obtained with $m_b=0$ (using MCFM), both at LO and at NLO. Using a
non-zero value of $m_b$ is not expected to have any impact on the
scale dependence of the result\footnote{Note that we always use
$m_b=4.62$~GeV in the determination of the scales in terms of
$\mu_0=m_b+M_W/2$ even in the results obtained with $m_b=0$.} and,
indeed, the scale dependence of the LO and NLO pair of curves is very
similar, with a shift due to the bottom-quark mass effects. 

While the
LO cross-section still has a 40\% uncertainty due to scale dependence,
this uncertainty is reduced at NLO to about 20\% for the
\emph{inclusive} and to about 10\% for the \emph{exclusive}
cross-section respectively. The uncertainties have been estimated
as the positive/negative deviation with respect to the mid-point of
the bands plotted in Fig.~\ref{fig:mu_dependence_band}, where each
band range is defined by the minimum and maximum value in the band.
We notice incidentally that the difference due to finite bottom-quark
mass effects is less significant than the theoretical uncertainty due
to the residual scale dependence in the
\emph{inclusive} case, but is comparable in size in the
\emph{exclusive} case.  Indeed, the finite bottom-quark mass effects
amount to about 8\% in both \emph{inclusive} and \emph{exclusive}
cases.

In Fig.~\ref{fig:sigma_ratio_NLO} we show the rescaled difference between
the total cross-sections obtained from our calculation (with $m_b\ne
0$) and with MCFM (with $m_b=0$) defined as follows:
\[\Delta\sigma=\sigma^{NLO}(m_b\ne 0)-\sigma^{NLO}(m_b=0) 
\; \frac{\sigma^{LO}(m_b\ne 0)}{\sigma^{LO}(m_b=0)} \; .\] 
As can be seen, within the statistical errors of the Monte Carlo
integration, the finite bottom-quark mass effects on the total
cross-sections at NLO are well described by the corresponding effects
at LO.

\begin{figure}[htb]
\begin{center}
\includegraphics[scale=0.6,angle=-90]{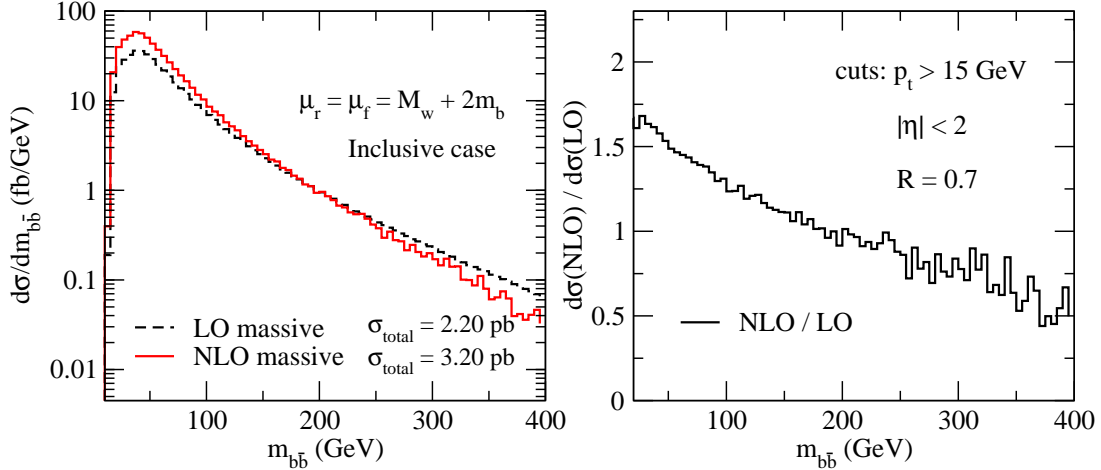} 
\caption[]{The \emph{inclusive} distribution $d\sigma/dm_{b\bar{b}}$
  in LO and NLO QCD. The right hand side plot shows the ratio of
  the LO and NLO distributions.}
\label{fig:mbb_dist_LO_vs_NLO_inc}
\end{center}
\end{figure}
\begin{figure}[htb]
\begin{center}
\includegraphics[scale=0.6,angle=-90]{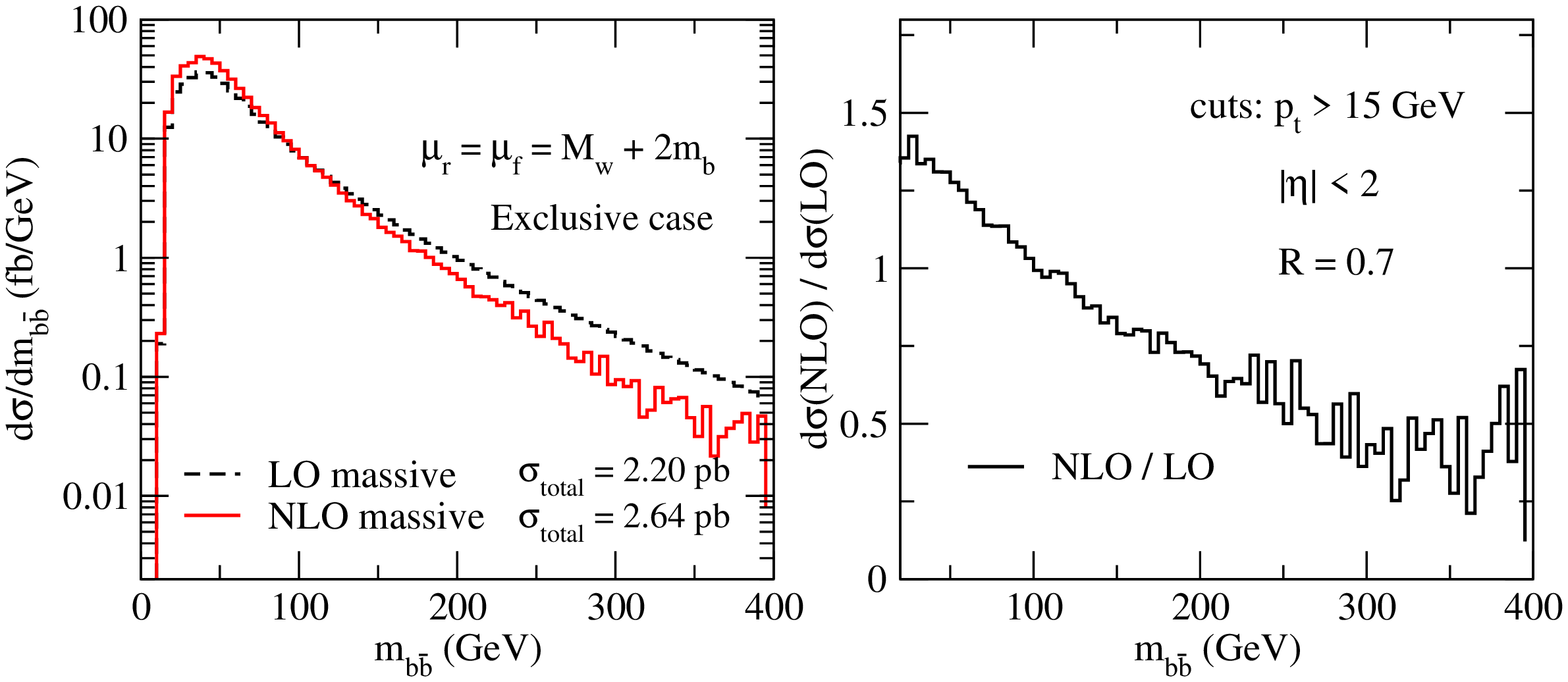} 
\caption[]{The \emph{exclusive} distribution $d\sigma/dm_{b\bar{b}}$
  in LO and NLO QCD. The right hand side plot shows the ratio of
  the LO and NLO distributions.}
\label{fig:mbb_dist_LO_vs_NLO_exc}
\end{center}
\end{figure}
\begin{figure}[htb]
\begin{center}
\includegraphics[scale=0.6,angle=-90]{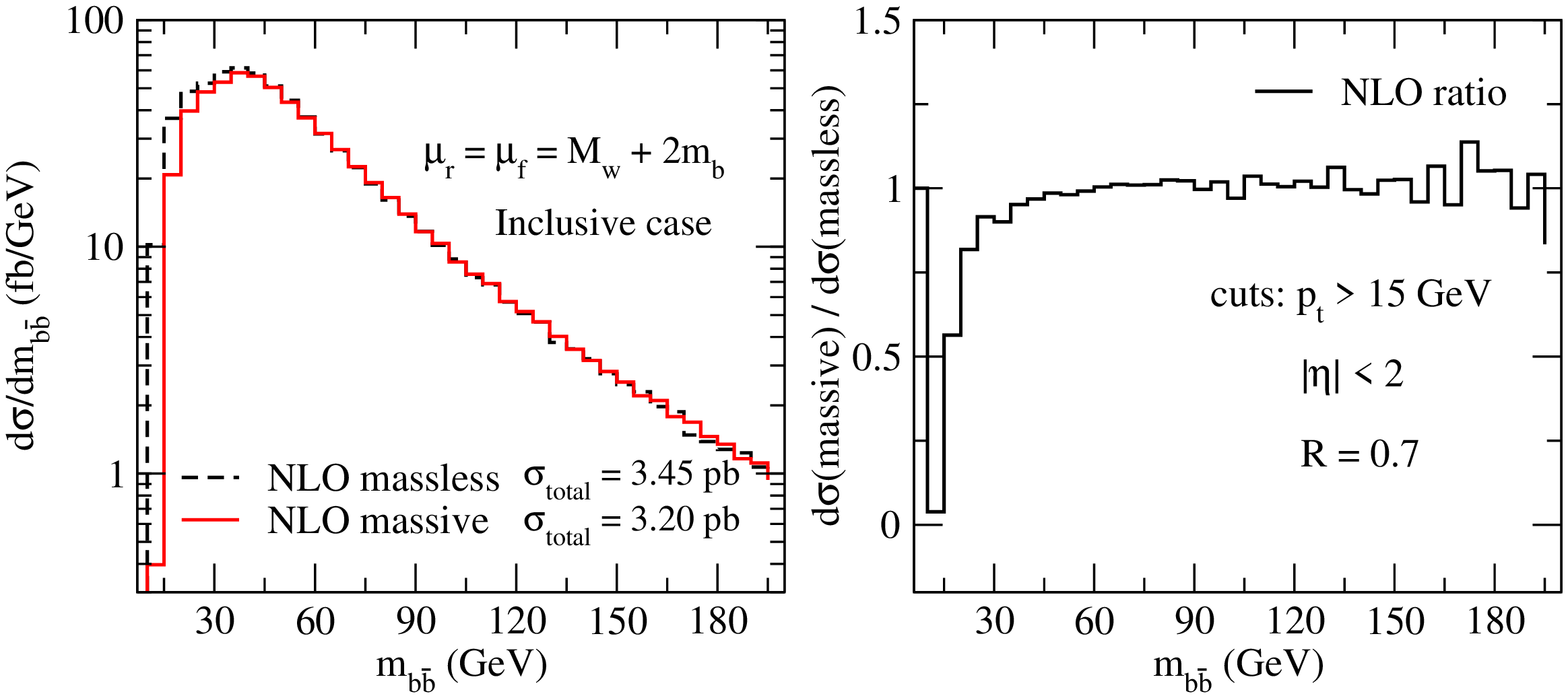} 
\caption[]{The \emph{inclusive} distribution $d\sigma/dm_{b\bar{b}}$
derived from our calculation (with $m_b\ne 0$) and from MCFM (with
$m_b=0$).  The right hand side plot shows the ratio of the two
distributions, $d\sigma(m_b\neq 0)/d\sigma(m_b=0)$.}
\label{fig:mbb_dist_NLO_inc}
\end{center}
\end{figure}
\begin{figure}[htb]
\begin{center}
\includegraphics[scale=0.6,angle=-90]{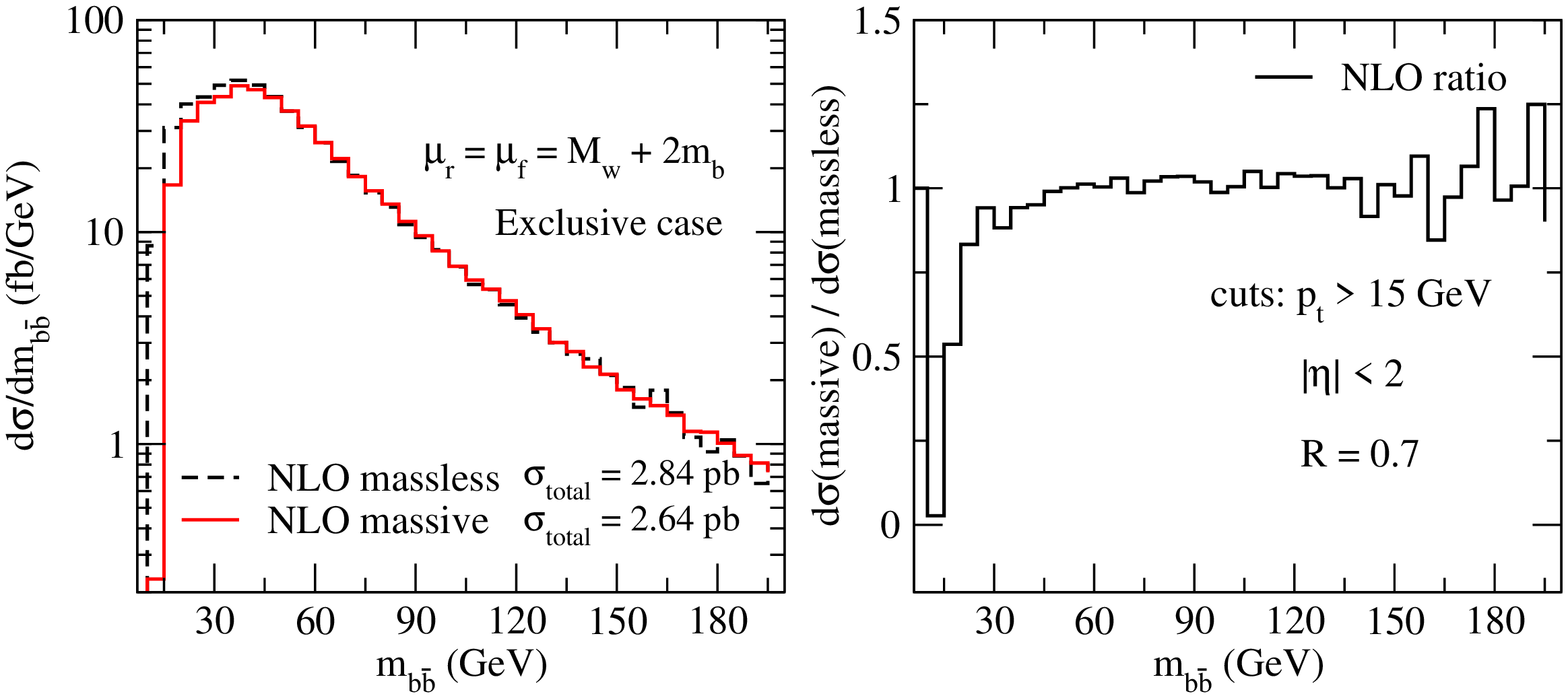} 
\caption[]{The \emph{exclusive} distribution $d\sigma/dm_{b\bar{b}}$
derived from our calculation (with $m_b\ne 0$) and from MCFM (with
$m_b=0$).  The right hand side plot shows the ratio of the two
distributions, $d\sigma(m_b\neq 0)/d\sigma(m_b=0)$.}
\label{fig:mbb_dist_NLO_exc}
\end{center}
\end{figure}
\begin{figure}[htb]
\begin{center}
\includegraphics[scale=0.6,angle=-90]{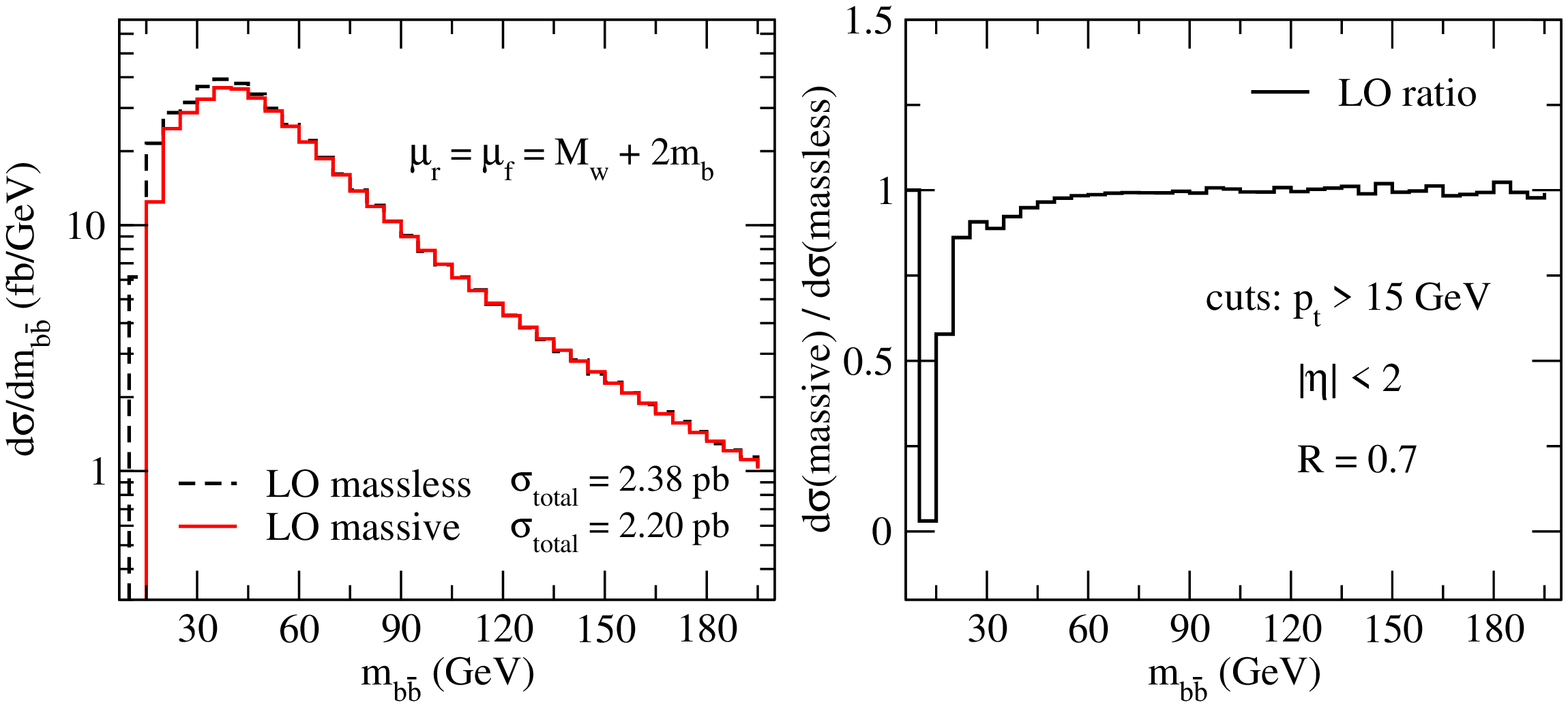} 
\caption[]{The LO distribution $d\sigma/dm_{b\bar{b}}$
derived from our calculation (with $m_b\ne 0$) and from MCFM (with
$m_b=0$). The right hand side plot shows the ratio of the two
distributions, $d\sigma(m_b\neq 0)/d\sigma(m_b=0)$.}
\label{fig:mbb_dist_LO}
\end{center}
\end{figure}
\begin{figure}[htb]
\begin{center}
\includegraphics*[scale=0.5]{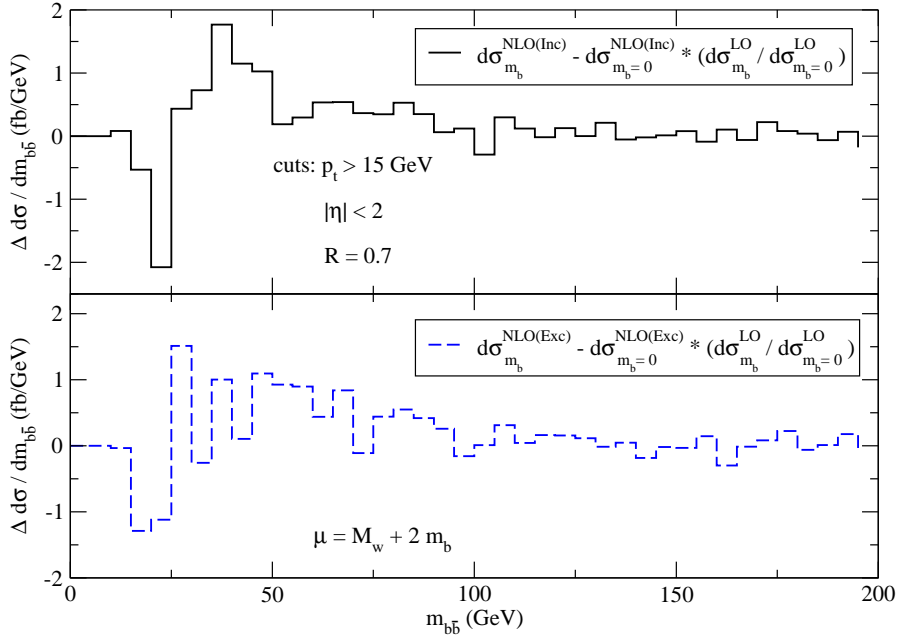} 
\caption[]{The $m_{b\bar b}$ distribution of the rescaled difference 
between our NLO calculation (with $m_b\ne 0$) and MCFM (with
$m_b=0$) for the \emph{inclusive} (upper plot) and \emph{exclusive} case (lower plot).}
\label{fig:mbb_ratio_NLO}
\end{center}
\end{figure}

Finally, in
Figs.~\ref{fig:mbb_dist_LO_vs_NLO_inc}-\ref{fig:mbb_dist_LO} we study
the distribution $d\sigma/dm_{b\bar{b}}$, where $m_{b\bar{b}}$ is the
invariant mass of the $b\bar{b}$ jet pair. The impact of NLO QCD
corrections on this distribution is illustrated in
Figs.~\ref{fig:mbb_dist_LO_vs_NLO_inc} and
\ref{fig:mbb_dist_LO_vs_NLO_exc} for the \emph{inclusive} and
\emph{exclusive} case respectively. We see that the NLO QCD
corrections affects the cross section quite substantially in
particular for low values of $m_{b\bar{b}}$. In each figure the right
hand side plot gives the ratio of the NLO and LO distributions,
providing a sort of $K$-factor bin by
bin. Figs.~\ref{fig:mbb_dist_NLO_inc} and \ref{fig:mbb_dist_NLO_exc}
compare the NLO $d\sigma/dm_{b\bar{b}}$ distributions obtained from
the massive and massless bottom-quark calculations. The results with
$m_b=0$ have been obtained using MCFM.  As expected, most of the
difference between the massless and massive bottom-quark
cross-sections is coming from the region of low invariant mass
$m_{b\bar{b}}$, both at LO and at NLO, where the cross-sections for
$m_b\ne 0$ are consistently below the ones with $m_b=0$.  For
completeness, we also show in Fig.~\ref{fig:mbb_dist_LO} the
comparison between massive ($m_b\neq 0$) and massless ($m_b=0$)
calculations at LO in QCD. The LO $m_{b\bar{b}}$ distribution for
massive bottom-quarks has been obtained both from our calculation and
from MCFM, which implements the $m_b\neq 0$ option at tree level, and
both results have been found in perfect agreement. As can be seen by
comparing Figs.~\ref{fig:mbb_dist_NLO_inc}-\ref{fig:mbb_dist_NLO_exc}
and Fig.~\ref{fig:mbb_dist_LO}, the impact of a non-zero bottom-quark
mass is almost not affected by including NLO QCD corrections.  
To illustrate this in more detail we show in
Fig.~\ref{fig:mbb_ratio_NLO} the rescaled difference between the $m_{b\bar
b}$ distributions obtained with our NLO calculation (with $m_b\ne 0$)
and with MCFM (with $m_b=0$) defined as follows:
\[\Delta \frac{d\sigma}{d m_{b \bar b}}=\frac{d\sigma^{NLO}}{d m_{b \bar b}}(m_b\ne 0)-\frac{d\sigma^{NLO}}{d m_{b \bar b}} (m_b=0)\; \frac{d\sigma^{LO}(m_b\ne 0)}{d\sigma^{LO}(m_b=0)} \; .\]
Apart from small deviations in the $m_{b\bar b}$ region below about
100 GeV, the finite bottom-quark mass effects at NLO are well
described by the LO calculations.

\section{Conclusions}
\label{sec:conclusions}
We have calculated the NLO QCD corrections to $q \bar q' \to
Wb\bar{b}$ production including full bottom-quark mass effects.  We
have presented numerical results for the total cross-section and the
invariant mass distribution of the bottom-quark jet pair
($m_{b\bar{b}}$) at the Tevatron for both massless and massive bottom
quarks. We apply the $k_T$ jet algorithm, require two $b$-tagged jets,
and impose kinematical cuts that are inspired by the D$\emptyset$ and
CDF searches for the SM Higgs boson in $WH$ production. The
bottom-quark mass effects amount to about 8\% of the total NLO QCD
cross-section and can impact the shape of the $m_{b\bar b}$
distributions, in particular in regions of low $m_{b\bar b}$. This is
relevant to SM Higgs searches in $WH$ associated production and to
searches for single-top production.

\bibliography{wbb_nlo}
\end{document}